\documentclass[10pt]{article}

\usepackage[margin=0.75in]{geometry}
\usepackage{microtype}
\usepackage{amsmath,amssymb,amsfonts}
\usepackage{booktabs}
\usepackage{graphicx}
\usepackage{url}
\usepackage[hidelinks]{hyperref}
\usepackage{listings}
\usepackage{enumitem}
\usepackage{xspace}
\usepackage{float}

\newcommand{\MS}{\textsc{MerkleSpeech}\xspace}
\newcommand{\MSv}{\textsf{MSv1}\xspace}
\newcommand{\CID}{\textsf{CID}}
\newcommand{\rid}{\textsf{rid}}
\newcommand{\kid}{\textsf{kid}}
\newcommand{\params}{\textsf{params}}
\newcommand{\paramshash}{\textsf{params\_hash}}

\title{\MS: Public-Key Verifiable, Chunk-Localised Speech Provenance via Perceptual Fingerprints and Merkle Commitments}

\author{
Tatsunori Ono$^{1}$ \quad\\
$^{1}$ Department of Computer Science, University of Warwick\\
Institutional: \texttt{tatsunori.ono@warwick.ac.uk}; Personal: \texttt{tatsunori.ono@outlook.com}
}

\date{}

\begin{document}
\maketitle

\begin{abstract}
Speech provenance in practice goes beyond detecting whether a watermark is present. Real workflows involve splicing, quoting, trimming, and platform-level transforms that may preserve some regions while altering others. Neural watermarking systems have made strides in robustness and localised detection~\cite{defossez2024audioseal}, but most deployments still produce detector outputs with no third-party verifiable cryptographic proof tying a specific time segment to an issuer-signed original. Provenance standards like C2PA do adopt signed manifests and Merkle-based fragment validation~\cite{c2pa2024spec23}, yet their bindings target encoded assets and tend to break under re-encoding or routine distribution processing.

We propose \MS, a system specification for \emph{public-key verifiable}, \emph{chunk-localised} speech provenance offering two tiers of assurance. The first, a \emph{robust watermark attribution layer} (WM-only), survives common distribution transforms---resampling, bandpass filtering, moderate noise---and answers ``was this chunk issued by a known party?''. The second, a \emph{strict cryptographic integrity layer} (\MSv), additionally verifies Merkle inclusion of the chunk's fingerprint under an issuer signature, answering ``does this chunk's fingerprint match the enrolled commitment?''. The system computes perceptual fingerprints over short speech chunks, commits them in a Merkle tree whose root is signed with an issuer key, and embeds a compact in-band watermark payload carrying a random content identifier and chunk metadata sufficient to retrieve Merkle inclusion proofs from a repository. Once the payload is extracted (via a public decoder or an issuer-provided service), all subsequent verification steps---signature check, fingerprint recomputation, Merkle inclusion---use only public information. The result is a splice-aware timeline indicating which regions pass each tier and why any given region fails. We describe the protocol, provide pseudocode, and present experiments targeting very low false positive rates under resampling, bandpass filtering, and additive noise, informed by recent audits that identify neural codecs as a major stressor for post-hoc audio watermarks~\cite{ozer2025assessmentneuralcodecs, oreilly2025deepwatermarksshallow, liu2024benchmarkingrobustness}.
\end{abstract}

\section{Introduction}
Advances in speech generation and editing have expanded both beneficial applications and the risk of voice-based fraud and misinformation. Proactive audio watermarking has become a useful complement to post-hoc forensic detection; modern learned systems offer improved robustness and, in some cases, localised detection maps~\cite{defossez2024audioseal, chen2023wavmark, li2024ideaw, singh2024silentcipher, oreilly2024maskmark}. But provenance needs often go beyond a binary watermark decision. Audio in the wild gets cropped, concatenated, partially quoted, and spliced, and many platforms apply resampling or re-encoding along the way. Recent benchmarks show that robustness can degrade sharply under certain transformation pipelines, with neural codecs flagged as an especially difficult regime~\cite{liu2024benchmarkingrobustness, oreilly2025deepwatermarksshallow, ozer2025assessmentneuralcodecs}.

Meanwhile, provenance initiatives have converged on signed manifests and cryptographic hashing---C2PA Content Credentials being the most prominent example~\cite{c2pa2024spec23, c2pa2023spec12}. These provide strong integrity for a particular encoded asset and include Merkle-based fragment validation in some configurations~\cite{c2pa2024spec23}. The trouble is that file- or container-level hashes generally break under benign transforms like transcoding or resampling, unless provenance is re-issued~\cite{c2pa2024spec23, simmons2024broadcastprovenance}. This gap motivates in-band approaches that remain meaningful after typical processing.

The question driving this paper is: \emph{can we build a practical system that produces cryptographically defensible, chunk-local provenance claims for speech, survives distribution transforms, and supports splice-aware verification?}

\paragraph{Approach.}
\MS combines four components: (i)~deterministic chunk fingerprints; (ii)~a Merkle commitment over chunk digests; (iii)~an issuer signature over the root; and (iv)~an in-band watermark payload carrying a compact pointer that lets any verifier retrieve proofs and check any chunk using only public information. Two verification tiers fall out of this design: \emph{WM-only} (watermark payload recovery) provides robust attribution that survives common distribution transforms, while \emph{\MSv} (full Merkle inclusion plus signature) gives strict integrity assurance that flags any post-enrollment modification. The gap between the two tiers is itself informative: a chunk that passes WM-only but fails \MSv has been altered since enrollment.

\paragraph{Contributions.}
\begin{itemize}[leftmargin=*]
  \item \textbf{Public-key verifiable, chunk-local speech provenance.} A protocol authenticating individual time chunks via Merkle inclusion proofs anchored by an issuer signature. The output is a splice-aware verification map over the audio timeline.
  \item \textbf{Pluggable perceptual commitment layer.} We describe two fingerprint options for speech chunks---SSL-embedding quantisation and a deterministic spectral baseline---together with a commitment construction that binds chunking parameters to prevent substitution attacks. The choice of fingerprint function governs the robustness--integrity trade-off; here we evaluate a strict MFCC baseline and discuss transform-tolerant alternatives.
  \item \textbf{Streaming-capable watermark channel and repository workflow.} A versioned payload format with ECC, paired with a repository design that avoids embedding large proofs in-band while still supporting third-party verification.
  \item \textbf{Rigorous evaluation.} Metrics and stress tests targeting extremely low false positive rates, localisation under splices, and robustness categories drawn from recent benchmarks and audits~\cite{liu2024benchmarkingrobustness, ozer2025assessmentneuralcodecs}.
\end{itemize}

\begin{figure}[H]
  \centering
  \includegraphics[width=0.98\textwidth]{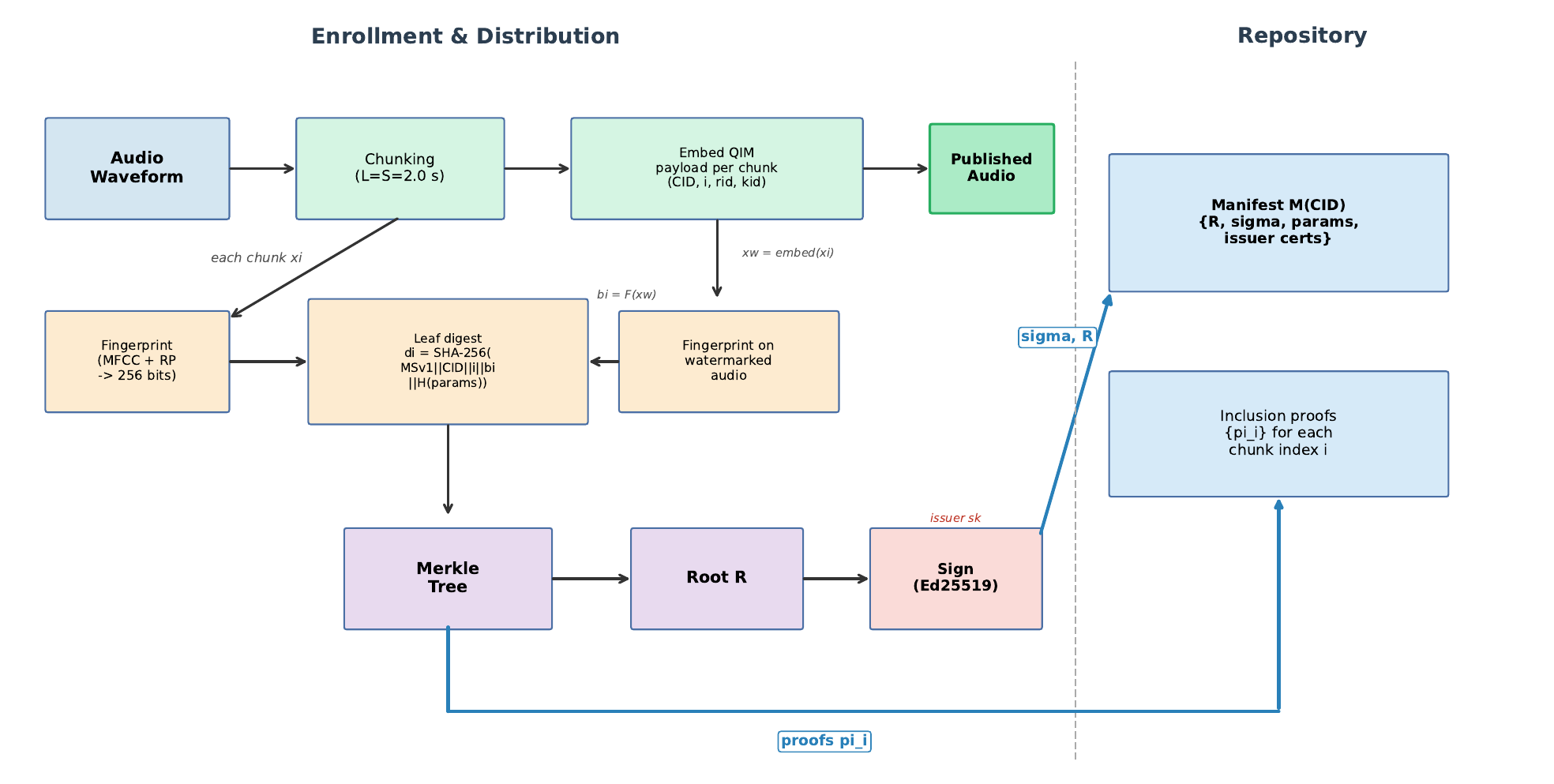}
  \caption{\textbf{\MS overview.} Enrollment: chunk audio, compute chunk fingerprints, build Merkle tree, sign root, publish manifest and inclusion proofs. Distribution: embed compact in-band payload per chunk. Verification: decode payload, fetch manifest and proofs, verify signature and Merkle inclusion per chunk, then aggregate to a splice-aware timeline.}
  \label{fig:system}
\end{figure}

\section{Related Work}

\paragraph{Neural post-hoc speech/audio watermarking.}
A growing body of work learns an embedder--detector pair, often trained end-to-end with differentiable augmentations. AudioSeal~\cite{defossez2024audioseal} introduced localised watermarking for proactive voice-cloning detection. WavMark~\cite{chen2023wavmark}, SilentCipher~\cite{singh2024silentcipher}, IDEAW~\cite{li2024ideaw}, and MaskMark~\cite{oreilly2024maskmark} explore various robustness--capacity trade-offs, while other lines of work target timbre-specific watermarking for cloning detection~\cite{liu2024timbrewatermark}, discrete intermediate representations~\cite{ji2025discrete}, cross-attention retrieval~\cite{xattnmark2025icml}, and key-enrichment strategies~\cite{xu2025wake}. What these systems share is a focus on imperceptibility, robustness, and localisation. What they generally lack is a public-key verifiable \emph{cryptographic proof of segment membership} tied to an issuer-signed commitment.

\paragraph{Robustness audits and benchmarks.}
Standardised benchmarking reveals wide variation in robustness across methods and attack types~\cite{liu2024benchmarkingrobustness}. In particular, audits have shown that learned enhancement pipelines and neural codecs can severely degrade post-hoc audio watermark performance~\cite{oreilly2025deepwatermarksshallow, ozer2025assessmentneuralcodecs}. These findings argue for treating neural codecs as first-class stress conditions rather than assuming robustness holds out of the box.

\paragraph{Fingerprint-based integrity verification.}
Closely related work on speech integrity via embedded fingerprints includes SpeeCheck~\cite{oreilly2026speecheck} and SpeechVerifier~\cite{yao2025speechverifier}. Classic audio fingerprinting \`{a} la Shazam~\cite{wang2003shazam} offers time-localisable descriptors robust to common distortions, and engineering baselines like Chromaprint~\cite{chromaprint2026} produce deterministic audio fingerprints. \MS shares their spirit but places greater emphasis on \emph{public-key verifiability} and \emph{repository-backed Merkle inclusion proofs}, aiming to support third-party provenance workflows that go beyond self-contained consistency checks.

\paragraph{Provenance standards and manifest approaches.}
C2PA defines signed manifests with hashing-based bindings and, in certain settings, Merkle-based fragment validation~\cite{c2pa2024spec23, c2pa2023spec12}. Work on broadcast provenance~\cite{simmons2024broadcastprovenance} combines open standards, metadata, watermarking, and cryptography, with discussion of piecewise validation. \MS complements these efforts by binding provenance to robust chunk fingerprints and carrying an in-band pointer, so that provenance can survive re-encoding or metadata stripping.

\paragraph{Publicly verifiable watermarking protocols.}
Puppy~\cite{isler2023puppy} explores how to transform symmetric watermark verification into publicly verifiable verification using secure hardware or secure computation. \MS targets a lighter-weight flow---signature verification plus Merkle inclusion proofs---paired with a robust watermark channel for discovery and indexing.

\section{Method: \MSv}

\subsection{Threat model}
\begin{center}
\fbox{
\begin{minipage}{0.97\linewidth}
\textbf{Threat model (v1).}
\begin{itemize}[leftmargin=*]
  \item \textbf{Verifier capabilities:} access to suspect audio; access to issuer public keys or certificates; online or cached access to a repository serving signed manifests and Merkle inclusion proofs; and access to a watermark decoder (a publicly released detector binary or a decoding service provided by the issuer). Note that watermark \emph{decoding} in our prototype relies on a secret permutation key. The ``public-key verifiable'' property applies to the Merkle signature layer: once the payload has been decoded by whatever means, every subsequent verification step---signature check, fingerprint recomputation, Merkle inclusion---requires only public information.
  \item \textbf{Adversary capabilities (black-box):} the adversary may apply common distribution transforms (transcoding, resampling, additive noise, enhancement) and perform edits such as cropping, concatenation, and splicing. Robustness audits motivate treating learned transforms and neural codecs as important stressors~\cite{ozer2025assessmentneuralcodecs, oreilly2025deepwatermarksshallow}.
  \item \textbf{Key assumptions:} the adversary does \emph{not} possess embedding key or model weights in a white-box manner and cannot forge issuer signatures. White-box and adaptive attacks are discussed as limitations.
  \item \textbf{Security goal:} produce a splice-aware timeline labelling which chunks can be publicly verified as belonging to an issuer-signed commitment, while maintaining extremely low false positive rates.
\end{itemize}
\end{minipage}
}
\end{center}

\subsection{Notation}
\begin{table}[H]
\centering
\caption{Core notation for \MSv.}
\label{tab:notation}
\begin{tabular}{@{}ll@{}}
\toprule
Symbol & Meaning \\
\midrule
$\CID$ & Random per-asset content identifier \\
$i$ & Chunk index \\
$L,S$ & Chunk length, stride (seconds) \\
$x_i$ & Original chunk waveform \\
$y_i$ & Observed/suspect chunk waveform \\
$b_i$ & Robust fingerprint bitstring for chunk $i$ \\
$d_i$ & Leaf digest (hash of metadata and $b_i$) \\
$R$ & Merkle root over leaves $\{d_i\}$ \\
$\sigma$ & Issuer signature over $(R, \CID, \paramshash, \dots)$ \\
$\pi_i$ & Merkle inclusion proof for leaf $i$ \\
\bottomrule
\end{tabular}
\end{table}

\subsection{Chunking and alignment}
Audio is segmented into chunks of length $L$ seconds with stride $S$. Non-overlapping chunking ($S = L$) gives a clean partitioning; overlapping chunking ($S < L$) allows finer splice-boundary localisation at the cost of redundant fingerprint computation and per-chunk payloads. Our experiments use $L = S = 2.0$\,s (non-overlapping). Chunking parameters are committed via $\paramshash$ (Section~\ref{sec:commitment}) to prevent an adversary from claiming verification under a different chunking scheme.

\subsection{Chunk fingerprints}
\label{sec:fingerprint}
\MS requires a mapping $F(\cdot)$ from a chunk waveform to a bitstring that is \emph{deterministic} (identical output on identical input) and \emph{content-sensitive} (edits and splices change the fingerprint):
\[
b_i = F(x_i) \in \{0, 1\}^m.
\]
How much transform tolerance the system provides depends on the fingerprint function. A strict function like the MFCC-based one we evaluate here detects any spectral modification, giving a strong integrity check. A more tolerant function---SSL-embedding quantisation, for instance---could survive benign processing at the expense of weaker integrity guarantees. We evaluate the strict variant and discuss tolerant alternatives as future work (Section~\ref{sec:limitations}).

\paragraph{Option A: SSL embeddings with random-projection binarization.}
A chunk embedding is computed using a pretrained self-supervised speech model (e.g., wav2vec~2.0)~\cite{baevski2020wav2vec2}. Let $e_i \in \mathbb{R}^d$ be a pooled embedding over the chunk. A fixed random projection $P \in \mathbb{R}^{d \times m}$ is applied and the result binarized by sign:
\[
b_i = \mathbb{I}\left[P^\top e_i \ge 0\right].
\]
The projection seed and pooling method are recorded in $\params$ and bound through $\paramshash$.

\paragraph{Option B: deterministic spectral fingerprint baseline.}
As a non-neural engineering baseline, we include a Chromaprint-style spectral fingerprinting pathway~\cite{chromaprint2026}, drawing on the broader tradition of time-localisable audio fingerprints~\cite{wang2003shazam}. This option is attractive for lightweight verification, though discriminability and short-window stability require empirical validation.

\paragraph{Remark on guarantees.}
These fingerprints are \emph{perceptual} rather than collision-resistant in the cryptographic sense. \MS therefore provides cryptographic authenticity \emph{conditioned on} the fingerprint function's sensitivity; we flag this limitation explicitly in Section~\ref{sec:limitations}.

\subsection{Cryptographic commitment and manifest}
\label{sec:commitment}
For each chunk index $i$, a leaf digest is computed with a cryptographic hash $H$ (SHA-256 in our implementation):
\[
d_i = H\big(\MSv \,\|\, \CID \,\|\, i \,\|\, b_i \,\|\, \paramshash\big).
\]
A Merkle tree is built over $\{d_i\}_{i=0}^{N-1}$, yielding root $R$. Because Merkle trees support inclusion proofs of size $O(\log N)$ per leaf~\cite{merkle1987digitalsignature}, individual chunks can be verified efficiently. The issuer signs:
\[
\sigma = \mathrm{Sign}_{sk}\big(R \,\|\, \CID \,\|\, \paramshash \,\|\, \textsf{issuer\_meta}\big),
\]
using Ed25519 or ECDSA-P256; the signature scheme and public key distribution are recorded in the manifest.

\paragraph{Manifest and repository.}
A signed manifest $M(\CID)$ is stored in a repository:
\[
M(\CID) = \big(\CID, R, \sigma, \params, \textsf{issuer\_certs}, \textsf{metadata}\big),
\]
together with inclusion proofs $\pi_i$ (precomputed or generated on demand). This follows the signed-manifest philosophy of provenance ecosystems~\cite{c2pa2024spec23, simmons2024broadcastprovenance}, but shifts the content binding from fragile file hashes to perceptual chunk fingerprints.

\subsection{Watermark payload and coding}
\label{sec:payload}
Embedding full Merkle proofs in-band is impractical given typical watermark capacities. Instead, \MS embeds a compact payload that enables look-up of $M(\CID)$ and the relevant proof $\pi_i$.

\paragraph{Payload fields (v1 template).}
Each per-chunk payload contains:
\begin{itemize}[leftmargin=*]
  \item \textbf{version} (4 bits): protocol version.
  \item $\CID$ (96--128 bits): random per-asset identifier, privacy-preserving by design.
  \item $i$ (16--24 bits): chunk index or synchronisation token.
  \item $\rid$ (64 bits): truncated root pointer for repository lookup (optional but helpful).
  \item $\kid$ (16 bits): issuer key identifier / certificate hint.
\end{itemize}
An error-correcting code suitable for short blocks (BCH or Reed--Solomon) protects these fields, with interleaving across the chunk to mitigate burst errors.

\paragraph{Watermark channel (implementation choices).}
\MS is \emph{channel-agnostic}: any watermark scheme capable of carrying the above payload through the target distribution transforms can serve as the embedding layer. For this paper we instantiate the channel with a QIM-STFT watermark (Section~\ref{sec:impl}) as a transparent, reproducible baseline. Learned embedder--detector architectures trained with differentiable augmentations~\cite{defossez2024audioseal, chen2023wavmark, oreilly2024maskmark, singh2024silentcipher, li2024ideaw, xu2025wake} are a natural upgrade path that could improve robustness without changing the \MS verification protocol. Neural codec and learned-transform pipelines are treated as first-class robustness stressors in our evaluation~\cite{ozer2025assessmentneuralcodecs, oreilly2025deepwatermarksshallow}.

\subsection{Verification protocol}
\label{sec:verification}
Given suspect audio, verification proceeds per chunk and is streaming-capable:

\begin{enumerate}[leftmargin=*]
  \item \textbf{Decode payload:} run the watermark detector to recover candidate tuples $(\widehat{\CID}, \hat{i}, \widehat{\rid}, \widehat{\kid})$ with confidence.
  \item \textbf{Fetch manifest:} query the repository for $M(\widehat{\CID})$ (and/or $\widehat{\rid}$) to obtain $(R, \sigma, \params, \paramshash)$ and the issuer's keys or certificates.
  \item \textbf{Verify signature:} check $\mathrm{Verify}_{pk}(\sigma, R \| \widehat{\CID} \| \paramshash \| \textsf{issuer\_meta})$.
  \item \textbf{Recompute fingerprint:} compute $b'_i = F(y_{\hat{i}})$ under $\params$ and $d'_i = H(\MSv \| \widehat{\CID} \| \hat{i} \| b'_i \| \paramshash)$.
  \item \textbf{Verify inclusion:} retrieve $\pi_{\hat{i}}$ and check $\mathrm{MerkleVerify}(d'_i, \pi_{\hat{i}}, R)$.
\end{enumerate}

A chunk is marked \textbf{Verified} only if both the signature and inclusion checks pass. Otherwise it is \textbf{Unverified}, accompanied by a reason code: no payload decoded, manifest missing, bad signature, or inclusion failure. When windows overlap, results are aggregated into a splice-aware timeline.

\begin{figure}[H]
  \centering
  \includegraphics[width=0.98\textwidth]{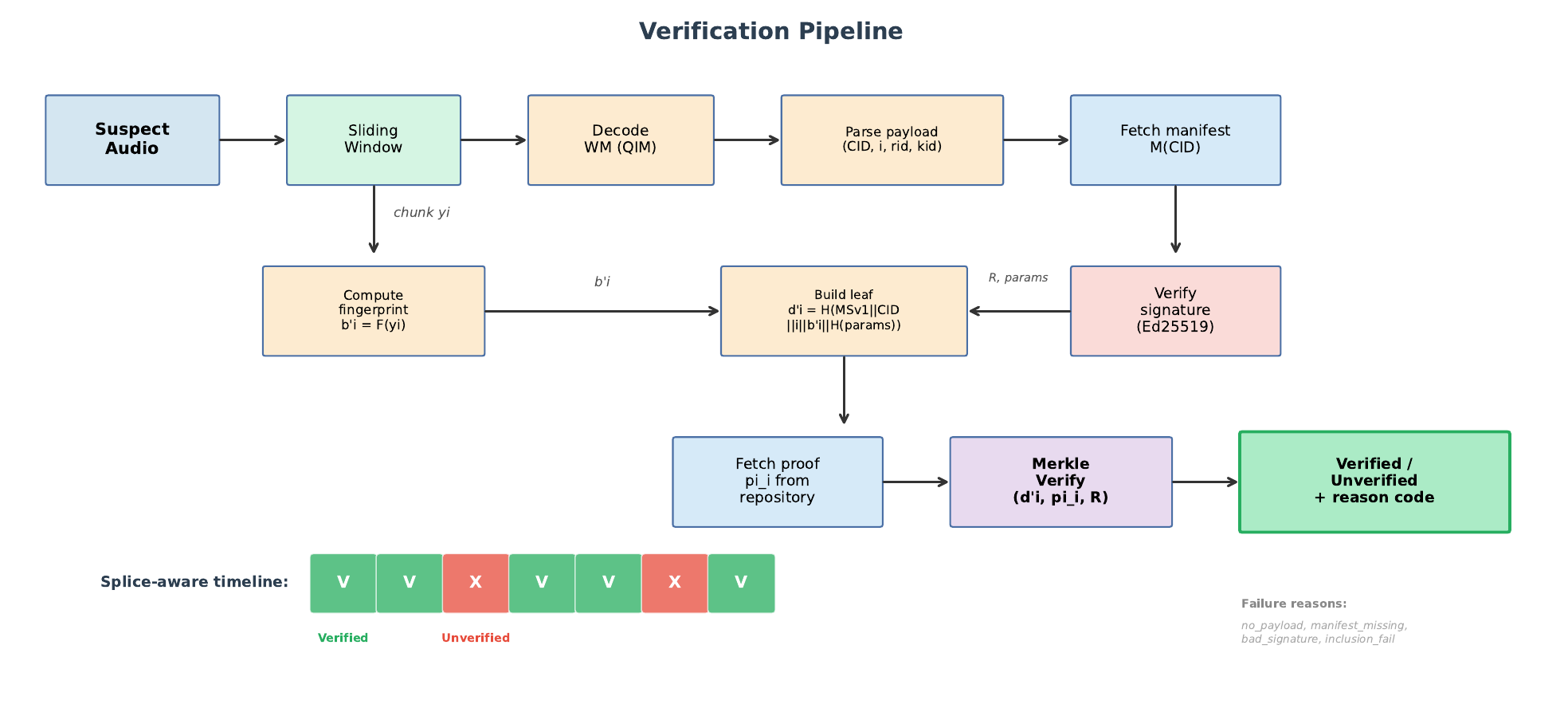}
  \caption{\textbf{Verification pipeline.} Streaming decode $\rightarrow$ manifest retrieval $\rightarrow$ signature verification $\rightarrow$ fingerprint recomputation and inclusion verification $\rightarrow$ splice-aware timeline with interpretable reason codes.}
  \label{fig:verify}
\end{figure}

\subsection{Algorithm / pseudocode (implementable interfaces)}
\label{sec:pseudocode}
The following pseudocode sketches the \MSv interfaces. It is meant to support reproducibility and does not provide guidance for watermark removal.

\begin{lstlisting}[language=Python,caption={Core APIs for \MSv enrollment and verification (pseudocode).},label={lst:apis}]
def enroll(audio_waveform, params, issuer_sk):
    # 1) choose random content identifier
    CID = random_bits(params.cid_bits)

    # 2) chunk audio into windows (length L, stride S)
    chunks = chunk(audio_waveform, L=params.L, S=params.S)

    # 3) compute robust fingerprints
    b = [Fingerprint(params.fingerprint_spec, x_i) for x_i in chunks]

    # 4) build leaf digests and Merkle tree
    params_hash = Hash(canonicalize(params))
    leaves = [Hash("MSv1" || CID || i || b_i || params_hash)
              for i, b_i in enumerate(b)]
    R, proofs = MerkleBuild(leaves)

    # 5) issuer signs root commitment
    sigma = Sign(issuer_sk, R || CID || params_hash || issuer_meta(params))

    # 6) publish manifest and proofs
    manifest = { "CID": CID, "R": R, "sigma": sigma,
                 "params": params, "params_hash": params_hash,
                 "issuer_meta": issuer_meta(params) }
    repo_put_manifest(CID, manifest)
    repo_put_proofs(CID, proofs)

    # 7) embed watermark payload per chunk (CID, index, rid, kid, version)
    watermarked = []
    for i, x_i in enumerate(chunks):
        payload = pack_payload(version=1, CID=CID, i=i,
                               rid=truncate64(R),
                               kid=params.kid)
        c = ECC_encode(payload, params.ecc)
        watermarked.append(Embedder(params.wm_model).embed(x_i, c))

    return overlap_add(watermarked, L=params.L, S=params.S)

def verify_streaming(suspect_audio, params, issuer_pk_resolver):
    out = []
    for (t, y_i) in sliding_windows(suspect_audio, L=params.L, S=params.S):
        decoded = Detector(params.wm_model).decode(y_i)
        if not decoded.decode_ok:
            out.append((t, "Unverified", "no_payload"))
            continue

        CID, i, rid, kid, version = parse_payload(decoded.payload)
        manifest = repo_get_manifest(CID, rid=rid)
        if manifest is None:
            out.append((t, "Unverified", "manifest_missing"))
            continue

        pk = issuer_pk_resolver(manifest["issuer_meta"], kid)
        if not Verify(pk, manifest["sigma"],
                      manifest["R"] || CID || manifest["params_hash"]
                      || manifest["issuer_meta"]):
            out.append((t, "Unverified", "bad_signature"))
            continue

        b_i = Fingerprint(manifest["params"]["fingerprint_spec"], y_i)
        d_i = Hash("MSv1" || CID || i || b_i || manifest["params_hash"])
        proof = repo_get_proof(CID, i)
        if proof is None:
            out.append((t, "Unverified", "proof_missing"))
            continue

        if not MerkleVerify(d_i, proof, manifest["R"]):
            out.append((t, "Unverified", "inclusion_fail"))
            continue

        out.append((t, "Verified", "ok"))

    return aggregate_timeline(out, stride=params.S)
\end{lstlisting}

\section{Experimental Setup}
\label{sec:experiments}

\subsection{Datasets}
All experiments use \textbf{LibriSpeech}~\cite{openslr12librispeech} (read English speech at 16\,kHz), specifically the \texttt{test-clean} subset (2\,620 files, roughly 5.4\,h). The first 200 files serve as the enrollment (positive) set; the remainder supply negatives for detection scoring, tail FPR estimation, and robustness evaluation. Audio is standardised to 16\,kHz, and the sampling rate is committed via $\paramshash$.

\subsection{Implementation details}
\label{sec:impl}

\paragraph{Watermark channel.}
The embedding layer is a QIM (Quantisation Index Modulation) watermark operating in the STFT log-magnitude domain with $\mathrm{n\_fft} = 1024$, $\mathrm{hop} = 256$, $\mathrm{win} = 1024$, embedding within the 300--3400\,Hz band. The default QIM step size is $\alpha = 0.6$, and each codeword bit is redundantly spread across up to 16 independent carrier bins selected by a secret permutation. Error correction uses a Reed--Solomon code RS(40,32), providing 8 bytes of parity. Embedding follows an iterative STFT projection loop: quantise the selected bins, synthesise via ISTFT, recompute the STFT from the resulting waveform, and re-quantise. The loop terminates once all bits decode correctly (typically 4--5 iterations), ensuring STFT consistency and preserving SNR.

\paragraph{Fingerprinting.}
Fingerprints are 256-bit MFCC-based descriptors: 13 MFCCs (40 mel bands) are computed, projected through a fixed random matrix, and binarised by sign. The projection matrix seed is committed in $\params$. Crucially, fingerprints are computed on the \emph{watermarked} audio at enrollment, so the Merkle inclusion proofs are valid for the published version of each chunk.

\paragraph{Cryptography.}
Leaf digests use SHA-256, issuer signatures use Ed25519, and the Merkle tree is binary with last-leaf duplication for non-power-of-two leaf counts. Payloads are packed in big-endian byte order into a 32-byte structure (version, CID, chunk index, truncated root, key ID).

\paragraph{Chunking.}
Chunks are $L = 2.0$\,s with stride $S = 2.0$\,s (non-overlapping), yielding 32\,000 samples per chunk at 16\,kHz.

\subsection{Baselines}
Rather than comparing against external systems (which do not natively support Merkle verification), we report two verification tiers derived from the same watermark channel:
\begin{itemize}[leftmargin=*]
  \item \textbf{WM-only (attribution):} the watermark payload RS-decodes successfully, the decoded content identifier resolves to a manifest in the repository, and the manifest's Ed25519 signature verifies under the issuer's public key. This answers ``was this chunk issued by a known, authenticated party?'' without requiring fingerprint recomputation or Merkle inclusion.
  \item \textbf{\MSv (strict integrity):} all WM-only checks pass, \emph{and} the chunk's recomputed fingerprint matches the enrolled Merkle leaf via an inclusion proof against the signed root. This answers ``does this chunk match the enrolled commitment exactly?''
\end{itemize}
Head-to-head comparison with neural watermarking systems such as AudioSeal~\cite{defossez2024audioseal}, WavMark~\cite{chen2023wavmark}, or SilentCipher~\cite{singh2024silentcipher} is deferred to future work.

\subsection{Metrics}
\begin{itemize}[leftmargin=*]
  \item \textbf{Screening:} ROC-AUC and FPR\textsubscript{score} at $10^{-4}$ and $10^{-6}$ for the repeat-consistency detector score (Section~\ref{sec:statrigor}).
  \item \textbf{Payload:} post-ECC bit error rate (BER) and full-message success rate.
  \item \textbf{Verification:} end-to-end \emph{verified rate} per chunk (WM-only: decode~+~signature; \MSv: decode~+~signature~+~fingerprint/Merkle), and FPR\textsubscript{verified} on held-out negatives.
  \item \textbf{Localisation:} chunk-level F1/IoU for splice regions.
  \item \textbf{Audio quality:} time-domain SNR between original and watermarked audio.
\end{itemize}

\subsection{Robustness suite}
The transform categories are parameterised and inspired by real-world conditions, following benchmark practice and recent audit findings~\cite{liu2024benchmarkingrobustness, ozer2025assessmentneuralcodecs, oreilly2025deepwatermarksshallow}:
\begin{itemize}[leftmargin=*]
  \item \textbf{Signal-level transforms:} additive white noise (SNR 10, 20, 30\,dB), resampling (8\,kHz and 12\,kHz round-trip), bandpass filtering (300--3400\,Hz), mild clipping (threshold 0.95), and reverberation (RT60~=~0.3\,s).
  \item \textbf{Splice scenarios:} segment insertion, segment removal/muting, and mixed-origin audio combining chunks from two different $\CID$s.
\end{itemize}

\subsection{Statistical rigor}
\label{sec:statrigor}
Tail FPR is evaluated using 3\,000\,000 negative windows drawn from the non-enrolled files (approximately 2\,420 files). Since only around 9\,700 non-overlapping 2\,s chunks fit in roughly 5.4\,h of audio, windows are obtained by \emph{random cropping with replacement}: for each file we sample random start offsets uniformly over valid positions and extract 2.0\,s windows, drawing up to 10\,000 windows per file. Because windows from the same file may overlap, they are not strictly independent; confidence intervals should be interpreted as approximate under this correlation structure.

The 3\,M windows are shuffled and split 50/50 into a validation set ($N_{\mathrm{val}} = 1.5 \times 10^6$) and a test set ($N_{\mathrm{test}} = 1.5 \times 10^6$). Two distinct false positive quantities are reported:
\begin{enumerate}[leftmargin=*]
  \item \textbf{FPR\textsubscript{score}} (screening): based on the repeat-consistency score from QIM carrier bins (Section~\ref{sec:impl}), calibrated so that higher values indicate watermark presence. For each target operating point $\tau \in \{10^{-4}, 10^{-6}\}$, the detection threshold is set on the validation split to achieve FPR${}_{\mathrm{val}} \le \tau$ (conservative), and FPR is then measured on the held-out test negatives. This score can serve as a compute-saving prefilter or stand-alone detector, but it is \emph{not} the WM-only or \MSv verification decision.
  \item \textbf{FPR\textsubscript{verified}} (pipeline): a negative window counts as a false positive only if the full verification pipeline returns \textbf{Verified}---meaning RS decode succeeds, the decoded CID resolves to a manifest, and all downstream checks pass. Evaluated on the same $N_{\mathrm{test}} = 1.5 \times 10^6$ held-out negatives.
\end{enumerate}
When false positives are observed, confidence intervals come from a binomial bootstrap (2\,000 resamples); when none are observed, we report an exact Clopper--Pearson 95\% upper bound. The robustness suite uses 90 positive chunks per transform condition.

\section{Results}
\label{sec:results}

All results are measured on LibriSpeech \texttt{test-clean} (200 enrolled files, remainder for negatives) with $\alpha = 0.6$, RS(40,32) ECC, and 2.0\,s non-overlapping chunks at 16\,kHz.

\subsection{Detection and payload accuracy}

Table~\ref{tab:benign} summarises detection and verification on clean (unperturbed) watermarked audio. Two distinct false positive metrics appear:
\begin{itemize}[leftmargin=*]
  \item \textbf{FPR\textsubscript{score}}: false positive rate of the repeat-consistency \emph{screening score}, thresholded on a validation split (Section~\ref{sec:statrigor}). Useful as a prefilter but \emph{not} the verification decision.
  \item \textbf{FPR\textsubscript{verified}}: false positive rate of the full pipeline (RS decode~$\to$~manifest lookup~$\to$~signature check; for \MSv also fingerprint/Merkle inclusion). This is the operationally relevant number.
\end{itemize}
The AUC\textsubscript{score} of 0.754 reflects the repeat-consistency screening score. Our focus is the low-FPR regime rather than maximising AUC; the operational decision rests on RS decode success followed by cryptographic checks. Both tiers achieve a 99.9\% decode rate and message match (694 of 695 chunks) with zero post-ECC bit errors. No end-to-end verified false positives were observed in $1.5 \times 10^6$ held-out negatives for either tier. For a negative to verify, it would need to both RS-decode successfully \emph{and} produce a CID resolving to a valid issuer-signed manifest (with MSv1 additionally requiring Merkle inclusion), which is vanishingly unlikely by construction.

On clean watermarked audio, both tiers reach a verified rate of 0.999. This confirms that fingerprints computed on the watermarked waveform at enrollment are deterministically reproduced at verification time, and that the Merkle proofs are valid for essentially all enrolled chunks. The tiers only diverge once transforms enter the picture (Section~\ref{sec:robustness}).

\begin{table*}[t]
\centering
\small
\caption{Benign evaluation on LibriSpeech. \textbf{Left:} detector screening score (repeat-consistency); FPR\textsubscript{score} is thresholded on a validation split and measured on 1.5\,M held-out negatives. \textbf{Right:} end-to-end verification pipeline; FPR\textsubscript{verified} counts negatives that RS-decode a valid payload \emph{and} pass all downstream checks. FPR@$10^{-6}$ is an upper confidence bound (Clopper--Pearson).}
\label{tab:benign}
\begin{tabular}{lcccccccccc}
\toprule
& \multicolumn{3}{c}{Screening score} & \multicolumn{6}{c}{End-to-end verification} \\
\cmidrule(lr){2-4} \cmidrule(lr){5-10}
Tier & AUC (screening) & FPR\textsubscript{score}@$10^{-4}$ & FPR\textsubscript{score}@$10^{-6}$ & BER & Decode & Msg match & Verified & FPR\textsubscript{verified} & $N_{\mathrm{pos}}$ \\
\midrule
WM-only & 0.754 & $8.3 \times 10^{-5}$ & $< 2 \times 10^{-6}$ & 0.000 & 0.999 & 0.999 & 0.999 & $0/1.5\mathrm{M}$ & 695 \\
\MSv & 0.754 & $8.3 \times 10^{-5}$ & $< 2 \times 10^{-6}$ & 0.000 & 0.999 & 0.999 & 0.999 & $0/1.5\mathrm{M}$ & 695 \\
\bottomrule
\end{tabular}
\end{table*}

\subsection{Robustness}
\label{sec:robustness}

\begin{figure}[H]
  \centering
  \includegraphics[width=\linewidth]{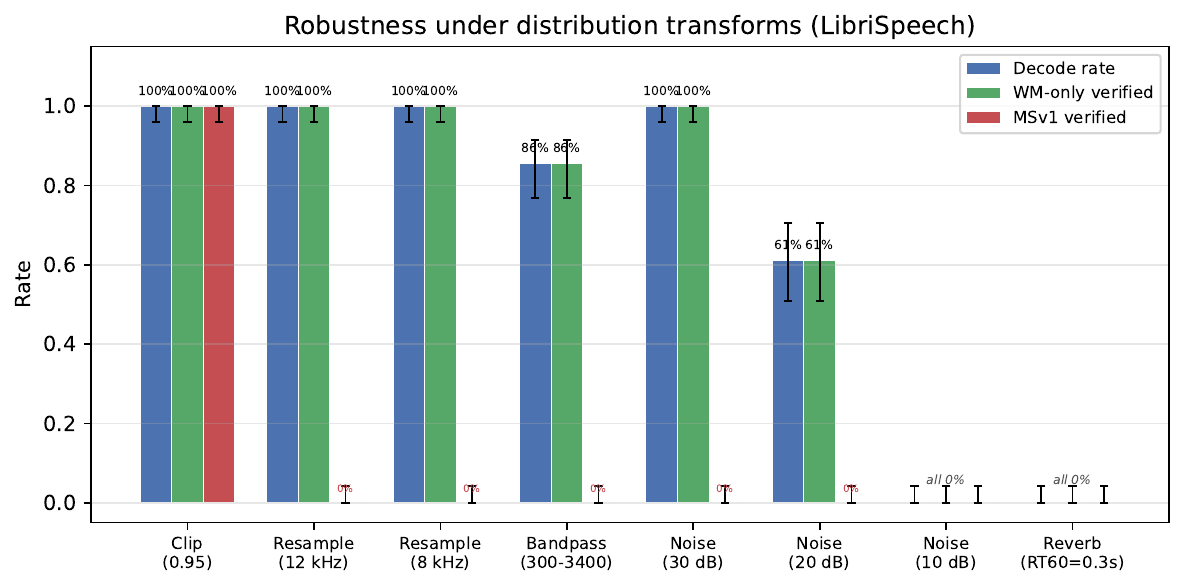}
  \caption{\textbf{Robustness under distribution transforms.} Grouped bars show decode rate, WM-only verification rate, and \MSv verification rate across eight transform conditions, ordered from mildest (left) to harshest (right). Error bars show Wilson 95\% binomial confidence intervals ($n = 90$ chunks per condition). The WM-only channel survives clipping, resampling, bandpass, and moderate noise; \MSv passes only for clipping (near-identity transform). The strict MFCC fingerprint makes \MSv deliberately sensitive to transforms in this instantiation; a transform-tolerant fingerprint (e.g., SSL embeddings) is future work (Section~\ref{sec:limitations}).}
  \label{fig:robustness}
\end{figure}

Figure~\ref{fig:robustness} presents verification performance across the robustness suite. To be precise about definitions: a chunk counts as \textbf{WM-only verified} when (1)~its watermark payload RS-decodes, (2)~the decoded CID resolves to a manifest, and (3)~the manifest's Ed25519 signature verifies. No detector-score threshold is involved; the decision boundary is RS decode success. Because all decoded payloads in our setup correspond to validly signed enrolled manifests, the \textbf{decode rate} and \textbf{WM-only verified rate} coincide.

The watermark channel proves quite robust to mild transforms: clipping (100\%), resampling at both 8 and 12\,kHz (100\%), noise at 30\,dB (100\%), and bandpass filtering (86\%). Under stronger perturbations, performance drops: 61\% at 20\,dB noise, 0\% at 10\,dB noise. Reverberation at RT60~=~0.3\,s also causes complete failure, which suggests the QIM embedding in log-magnitude is vulnerable to temporal smearing.

The \MSv verified rate, meanwhile, is 0\% under every transform except clipping (100\%). This is by design: any transform that alters the STFT magnitude will change the MFCC fingerprint, which in turn causes the Merkle inclusion proof to fail. The whole point of the integrity layer is to \emph{detect} post-distribution modifications, not to be invariant to them. The practical upshot is that the gap between the two tiers carries diagnostic meaning. When a chunk passes WM-only but fails \MSv, the natural reading is that the chunk was watermarked by this issuer but has undergone some modification since enrollment.

\subsection{Quality--robustness trade-off}

\begin{figure}[H]
  \centering
  \includegraphics[width=\linewidth]{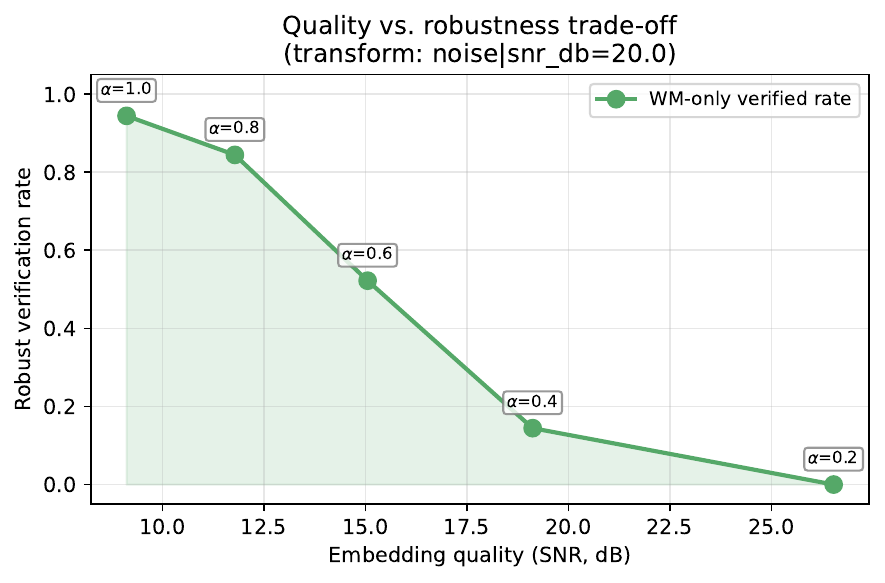}
  \caption{\textbf{Quality--robustness trade-off.} WM-only verification rate under additive noise (SNR~=~20\,dB) as a function of embedding quality (time-domain SNR), evaluated on the same 90-chunk subset as Figure~\ref{fig:robustness}. Each point uses an independent per-$\alpha$ enrollment (distinct CID and Merkle tree), so the $\alpha = 0.6$ rate may differ slightly from Figure~\ref{fig:robustness} due to payload-dependent QIM bit patterns; both fall within each other's 95\% CIs.}
  \label{fig:tradeoff}
\end{figure}

Figure~\ref{fig:tradeoff} traces the trade-off between embedding quality (SNR) and robustness as $\alpha$ varies over $\{0.2, 0.4, 0.6, 0.8, 1.0\}$, with additive noise at 20\,dB as the test condition. Increasing $\alpha$ widens the QIM step, improving robustness but lowering SNR. At $\alpha = 0.2$ the embedding is nearly imperceptible (SNR${}\approx 27$\,dB) but fragile (0\% verification under 20\,dB noise). At $\alpha = 1.0$ the SNR falls to 9.1\,dB but 94\% of chunks verify. The default $\alpha = 0.6$ strikes a reasonable middle ground: 15.1\,dB SNR with 52\% robust verification. (Each $\alpha$ point uses an independent enrollment with a distinct CID and Merkle tree, so payload-dependent QIM bit patterns vary; the $\alpha = 0.6$ rates in this sweep and in Figure~\ref{fig:robustness} fall within each other's 95\% binomial confidence intervals.)

\subsection{Splice detection}

Table~\ref{tab:splice} reports splice-aware localisation under \emph{boundary-aligned} edits: entire 2.0\,s chunks are inserted, removed, or substituted, with non-overlapping chunking ($L = S = 2.0$\,s). Both WM-only and \MSv achieve perfect IoU and macro-F1 under these conditions, because chunk-level payload decoding unambiguously identifies the source $\CID$ for each chunk. These perfect scores are expected by construction---boundary-aligned splicing reduces to per-chunk payload classification---and we include them mainly as a sanity check that chunk identity propagates correctly through the full pipeline.

\paragraph{Limitations of the current splice regime.}
Misaligned splices---inserting 0.5\,s mid-chunk, or cutting at an arbitrary sample offset---would disrupt both the watermark and fingerprint for the affected chunk but could not be localised below the 2.0\,s chunk granularity. This resolution limit is inherent to non-overlapping chunking. Overlapping chunking ($S < L$) would sharpen boundary localisation at the cost of redundant per-chunk payloads and fingerprint computation; we leave evaluation of such configurations to future work.

\begin{table}[H]
\centering
\small
\caption{Splice evaluation (\textbf{boundary-aligned}, $L = S = 2.0$\,s). All edits are aligned to chunk boundaries, so the task reduces to per-chunk payload classification---perfect scores are expected by construction. IoU is computed on a binary tamper mask; mixed-origin uses three-way macro-F1 over $\{\mathrm{CID}_A, \mathrm{CID}_B, \varnothing\}$. See text for discussion of misaligned splices.}
\label{tab:splice}
\begin{tabular}{lccc}
\toprule
Scenario & Metric & \MSv & WM-only \\
\midrule
Insert segment & IoU & 1.000 & 1.000 \\
Remove/mute segment & IoU & 1.000 & 1.000 \\
Mixed origin & F1$_{\mathrm{macro}}$ & 1.000 & 1.000 \\
\bottomrule
\end{tabular}
\end{table}

\section{Discussion}

The core design choice in \MS is to keep signal-layer robustness (watermark payload recovery) and cryptographic authenticity (signature plus Merkle inclusion) cleanly separated. The experimental results illustrate why this matters: under moderate noise (20\,dB), 61\% of chunks still pass WM-only verification while none pass \MSv, since the transform alters the MFCC fingerprint. Far from being a failure, this is exactly the diagnostic the system is intended to provide---the chunk was watermarked by a particular issuer, but its content has changed since enrollment.

Compared to detector-only watermarking, the Merkle inclusion proof and issuer signature give a third-party verifiable artifact grounded in public-key infrastructure, which aligns with the direction of provenance ecosystems built on signed manifests~\cite{c2pa2024spec23}. Relative to embedded-fingerprint integrity schemes like SpeeCheck~\cite{oreilly2026speecheck} and SpeechVerifier~\cite{yao2025speechverifier}, the advantage of the Merkle approach is scalability: per-chunk proofs are compact ($O(\log N)$) without requiring large in-band embeddings, and manifests can slot into existing provenance metadata workflows~\cite{simmons2024broadcastprovenance}.

\paragraph{Robustness--integrity trade-off.}
The experiments expose a genuine design tension. WM-only verification tolerates resampling, bandpass filtering, and moderate noise, while \MSv catches any spectral change but is correspondingly brittle to benign transforms. A practical deployment might use WM-only as a first-pass attribution check and reserve \MSv for contexts demanding strict integrity. Looking ahead, replacing the MFCC fingerprint with a transform-tolerant alternative---stabilised SSL embeddings, for instance---could allow \MSv to survive benign processing while still flagging adversarial edits, though such a fingerprint would inevitably trade off some integrity strength.

\paragraph{Splice detection.}
The perfect splice scores in Table~\ref{tab:splice} are a direct consequence of the per-chunk payload design: since each chunk independently carries its own $\CID$ and index, spliced regions from different sources are identified at chunk granularity. This is a structural property of the \MS protocol rather than something specific to the QIM watermark---any sufficiently robust watermark with per-chunk payload capacity would work.

\section{Limitations}
\label{sec:limitations}
\begin{itemize}[leftmargin=*]
  \item \textbf{Perceptual fingerprints are not collision-resistant.} The MFCC fingerprint is strict: it flags any spectral modification, making \MSv deliberately sensitive to transforms. A more tolerant fingerprint (e.g., SSL embeddings) could let \MSv survive benign processing but would weaken integrity guarantees. Either way, perceptual fingerprints can collide, so \MS authenticity is ultimately conditioned on the chosen fingerprint function (Section~\ref{sec:fingerprint}).
  \item \textbf{Neural codec robustness is not evaluated.} The robustness suite covers signal-level transforms---noise, resampling, bandpass, clipping, reverb---but omits neural codec conditions. Given that prior audits single out neural codecs as a particularly challenging regime for post-hoc watermarks~\cite{ozer2025assessmentneuralcodecs, oreilly2025deepwatermarksshallow}, evaluation under codecs is a clear next step.
  \item \textbf{Repository dependency (v1).} Public verification requires fetching manifests and proofs unless they are cached. This is a deliberate trade-off: keeping in-band payloads small.
  \item \textbf{White-box and adaptive attackers.} Full access to embedding/detector models or adaptive optimisation could erode robustness. Protocol-level defences (e.g., the publicly verifiable approach of Puppy~\cite{isler2023puppy}) are one avenue; we leave this to future work.
\end{itemize}

\section{Ethics \& Responsible Use}

Watermarking and provenance tools can serve accountability, but they can equally be misused for surveillance or coercive control. \MS is designed to minimise embedded metadata, the $\CID$ is random and carries no user or device identifiers, and to support transparent, auditable issuer key management. Our evaluation is described at the level of robustness categories and metrics; we deliberately avoid providing operational guidance aimed at watermark removal or evasion. Any real deployment should include governance mechanisms, opt-out and key rotation policies, and published false-positive audits, in line with concerns raised by robustness benchmarks~\cite{liu2024benchmarkingrobustness}.

\section{Conclusion}

This paper introduced \MS, a system for public-key verifiable, chunk-local speech provenance that is explicitly splice-aware. It offers two tiers of assurance: robust attribution via WM-only, which survives common distribution transforms, and strict cryptographic integrity via \MSv, which detects post-enrollment modifications. The architecture ties together deterministic chunk fingerprints, Merkle-tree commitments, issuer signatures, and an in-band watermark pointer for repository-backed proof retrieval and third-party verification. On LibriSpeech, the system achieves a 99.9\% payload decode rate on clean audio, robust WM-only recovery under resampling, bandpass filtering, and moderate noise, and perfect boundary-aligned splice localisation. The most pressing directions for future work are integrating transform-tolerant SSL fingerprints (to give \MSv a chance of surviving benign transforms), evaluating under neural codec stress conditions~\cite{ozer2025assessmentneuralcodecs, oreilly2025deepwatermarksshallow}, and moving to overlapping chunking for finer splice-boundary resolution.

\section*{Acknowledgements}
We thank Professor Gene Tsudik for his arXiv endorsement to submit papers in the subject class cs.CR (Cryptography and Security).


\appendix
\section{Reproducibility Checklist and Additional Implementation Details}
\label{app:repro}

\subsection{Reproducibility checklist (v1)}
\begin{itemize}[leftmargin=*]
  \item \textbf{Code availability:} Provide source for (i)~watermark embedder/detector, (ii)~fingerprint extraction, (iii)~Merkle build/verify, (iv)~signing/verification, (v)~evaluation harness.
  \item \textbf{Parameter binding:} Publish a canonical serialization of $\params$ (e.g., canonical JSON/CBOR) and define $\paramshash = H(\textsf{canonicalize}(\params))$.
  \item \textbf{Fingerprint spec:} Document fingerprint option (SSL+projection or deterministic baseline), embedding extractor version, pooling, projection seed, bit length $m$, and any stabilization heuristics.
  \item \textbf{Chunking:} Report $L$, $S$, boundary policy (pad/drop), and aggregation rule for overlapping verification maps.
  \item \textbf{Cryptography:} Report hash function, Merkle tree convention (leaf order, left/right concatenation), signature scheme, and key distribution model (cert chain / pinned keys).
  \item \textbf{Payload format:} Publish bit-level packing, ECC scheme/parameters, interleaving scheme, repetition schedule, and decoder confidence thresholds.
  \item \textbf{Datasets/licenses:} Record dataset versions and licensing sources~\cite{openslr12librispeech, vctkdatashare, commonvoicecriteria}.
  \item \textbf{Evaluation protocol:} Define negative set construction for low-FPR estimation; report window counts and CIs for tail metrics.
  \item \textbf{Transform suite:} Publish transform categories and severity grids used for defensive evaluation, including neural codec stress conditions~\cite{ozer2025assessmentneuralcodecs}.
  \item \textbf{Randomness/seeds:} Fix and report seeds for training, projection matrices, and evaluation sampling.
  \item \textbf{Hardware/software:} Report PyTorch/NumPy versions, CUDA/cuDNN, and hardware specs for speed metrics.
\end{itemize}

\subsection{Merkle tree conventions (recommended)}
To avoid ambiguity across implementations~\cite{merkle1987digitalsignature}, we recommend:
\begin{itemize}[leftmargin=*]
  \item Leaves are the byte strings $d_i$ in chunk-index order $i = 0, \dots, N-1$.
  \item Internal nodes hash the concatenation of left child then right child.
  \item If $N$ is not a power of two, either (a)~duplicate the last leaf in a level, or (b)~use a standard ``Merkle with padding'' rule; the chosen rule must be specified in $\params$ and committed via $\paramshash$.
\end{itemize}

\subsection{Repository API sketch (v1)}
A minimal repository can be implemented as a key-value store with:
\begin{itemize}[leftmargin=*]
  \item \texttt{GET /manifest/\{CID\}} $\rightarrow$ $(R, \sigma, \params, \paramshash, \textsf{issuer\_meta})$
  \item \texttt{GET /proof/\{CID\}/\{i\}} $\rightarrow$ Merkle proof $\pi_i$
\end{itemize}
Clients should cache manifests and proofs to reduce latency and mitigate availability risks.

\subsection{Experiment parameters}
\begin{itemize}[leftmargin=*]
  \item \textbf{Random seed:} all experiments use seed~=~123; the fingerprint projection matrix uses seed~=~1122 ($= 123 + 999$), and the watermark permutation key uses seed~=~1460 ($= 123 + 1337$).
  \item \textbf{Dataset:} LibriSpeech \texttt{test-clean} (2620 FLAC files, approximately 5.4 hours), 200 files for positive enrollment, remainder for negative scoring.
  \item \textbf{Negative windows:} 3\,000\,000 windows sampled for tail FPR estimation, split 50/50 into validation and test sets.
  \item \textbf{Robustness:} 90 positive chunks (30 files~$\times$~3 chunks) per transform condition.
  \item \textbf{Hardware \& software:} Apple M4 (ARM64), CPU only. Python~3.13.2, PyTorch~2.10.0, torchaudio~2.10.0, NumPy~2.4.2, PyNaCl~1.6.2 (Ed25519), reedsolo~1.7.0, soundfile~0.13.1, macOS~15. Full experiment runtime is approximately 60--90 minutes.
  \item \textbf{Code:} We plan to release the full source code (watermark, fingerprint, Merkle, evaluation harness) as an open-source repository. In the interim, the pseudocode in Section~\ref{sec:pseudocode} and the parameter details above are intended to enable independent reimplementation.
\end{itemize}

\subsection{Figure generation}
The result figures can be reproduced locally with:
\begin{center}
\verb|python scripts/make_results_figures.py --results results_full --fig_dir fig|
\end{center}
This script reads the JSON evaluation outputs and renders the robustness bar chart and quality--robustness trade-off curve as PDF figures.

\begin{thebibliography}{10}

\bibitem{defossez2024audioseal}
Alexandre Defossez et~al.
\newblock Proactive detection of voice cloning with localised watermarking.
\newblock \url{https://arxiv.org/abs/2401.17264}, 2024.
\newblock Accessed 2026-02-04.

\bibitem{c2pa2024spec23}
{Coalition for Content Provenance and Authenticity (C2PA)}.
\newblock Content credentials: C2PA technical specification (v2.3).
\newblock
  \url{https://c2pa.org/specifications/specifications/2.3/specs/C2PA_Specification.html},
  2024.
\newblock Accessed 2026-02-04.

\bibitem{ozer2025assessmentneuralcodecs}
Yigitcan Ozer et~al.
\newblock A comprehensive real-world assessment of audio watermarking
  algorithms: Will they survive neural codecs?
\newblock \url{https://arxiv.org/pdf/2505.19663}, 2025.
\newblock Accessed 2026-02-04.

\bibitem{oreilly2025deepwatermarksshallow}
Peter O'Reilly et~al.
\newblock Deep audio watermarks are shallow.
\newblock
  \url{https://oreillyp.github.io/assets/manuscript/deep_watermarks_shallow.pdf},
  2025.
\newblock Accessed 2026-02-04.

\bibitem{liu2024benchmarkingrobustness}
Hao Liu et~al.
\newblock Benchmarking robustness of audio watermarking.
\newblock
  \url{https://proceedings.neurips.cc/paper_files/paper/2024/file/5d9b7775296a641a1913ab6b4425d5e8-Paper-Datasets_and_Benchmarks_Track.pdf},
  2024.
\newblock Accessed 2026-02-04.

\bibitem{chen2023wavmark}
Guangyu Chen et~al.
\newblock Wavmark: Watermarking for audio generation.
\newblock \url{https://arxiv.org/abs/2308.12770}, 2023.
\newblock Accessed 2026-02-04.

\bibitem{li2024ideaw}
Pengfei Li et~al.
\newblock Ideaw: Robust neural audio watermarking with invertible neural
  networks.
\newblock \url{https://aclanthology.org/2024.emnlp-main.258.pdf}, 2024.
\newblock Accessed 2026-02-04.

\bibitem{singh2024silentcipher}
Mayank~Kumar Singh et~al.
\newblock Silentcipher: Deep audio watermarking.
\newblock
  \url{https://www.isca-archive.org/interspeech_2024/singh24_interspeech.pdf},
  2024.
\newblock Accessed 2026-02-04.

\bibitem{oreilly2024maskmark}
Peter O'Reilly et~al.
\newblock Maskmark: Robust neural watermarking for real and synthetic speech.
\newblock
  \url{https://interactiveaudiolab.github.io/assets/papers/oreilly_jin_su_pardo_watermark.pdf},
  2024.
\newblock Accessed 2026-02-04.

\bibitem{c2pa2023spec12}
{Coalition for Content Provenance and Authenticity (C2PA)}.
\newblock C2PA technical specification (v1.2).
\newblock
  \url{https://spec.c2pa.org/specifications/specifications/1.2/specs/C2PA_Specification.html},
  2023.
\newblock Accessed 2026-02-04.

\bibitem{simmons2024broadcastprovenance}
J.~C. Simmons et~al.
\newblock Interoperable provenance authentication of broadcast media using open
  standards-based metadata, watermarking and cryptography.
\newblock \url{https://arxiv.org/pdf/2405.12336}, 2024.
\newblock Accessed 2026-02-04.

\bibitem{liu2024timbrewatermark}
Chang Liu et~al.
\newblock Detecting voice cloning attacks via timbre watermarking.
\newblock
  \url{https://www.ndss-symposium.org/wp-content/uploads/2024-200-paper.pdf},
  2024.
\newblock Accessed 2026-02-04; NDSS 2024. (Also reported with
  arXiv:2312.03410).

\bibitem{ji2025discrete}
Shuai Ji et~al.
\newblock Speech watermarking with discrete intermediate representations.
\newblock \url{https://ojs.aaai.org/index.php/AAAI/article/view/34600/36755},
  2025.
\newblock Accessed 2026-02-04.

\bibitem{xattnmark2025icml}
{ICML 2025 Poster Page}.
\newblock Learning robust audio watermarking with cross-attention (xattnmark).
\newblock \url{https://icml.cc/virtual/2025/poster/43452}, 2025.
\newblock Accessed 2026-02-04.

\bibitem{xu2025wake}
Yifan Xu et~al.
\newblock Wake: Watermarking audio with key enrichment.
\newblock
  \url{https://www.isca-archive.org/interspeech_2025/xu25f_interspeech.pdf},
  2025.
\newblock Accessed 2026-02-04.

\bibitem{oreilly2026speecheck}
Peter O'Reilly et~al.
\newblock SpeeCheck: Self-contained speech integrity verification via embedded
  acoustic fingerprints.
\newblock
  \url{https://openreview.net/pdf/003ec7b327997a2dfb4c22ea0570eda8d565dce0.pdf},
  2026.
\newblock Accessed 2026-02-04.

\bibitem{yao2025speechverifier}
Lingfeng Yao et~al.
\newblock Speechverifier: Robust acoustic fingerprint against tampering attacks
  via watermarking.
\newblock \url{https://arxiv.org/abs/2505.23821}, 2025.
\newblock Accessed 2026-02-04.

\bibitem{wang2003shazam}
Avery Li-Chun Wang.
\newblock An industrial-strength audio search algorithm.
\newblock \url{https://www.ee.columbia.edu/~dpwe/papers/Wang03-shazam.pdf},
  2003.
\newblock Accessed 2026-02-04.

\bibitem{chromaprint2026}
{AcoustID}.
\newblock Chromaprint: An audio fingerprinting library.
\newblock \url{https://acoustid.org/chromaprint}, 2026.
\newblock Accessed 2026-02-04.

\bibitem{isler2023puppy}
Devris Isler et~al.
\newblock Puppy: A publicly verifiable watermarking protocol.
\newblock \url{https://arxiv.org/abs/2312.09125}, 2023.
\newblock Accessed 2026-02-04.

\bibitem{baevski2020wav2vec2}
Alexei Baevski et~al.
\newblock wav2vec 2.0: A framework for self-supervised learning of speech
  representations.
\newblock \url{https://arxiv.org/abs/2006.11477}, 2020.
\newblock Accessed 2026-02-04.

\bibitem{merkle1987digitalsignature}
Ralph Merkle.
\newblock A digital signature based on a conventional encryption function.
\newblock \url{https://link.springer.com/chapter/10.1007/3-540-48184-2_32},
  1987.
\newblock Accessed 2026-02-04.

\bibitem{openslr12librispeech}
{OpenSLR}.
\newblock Librispeech asr corpus (openslr 12).
\newblock \url{https://www.openslr.org/12}, 2026.
\newblock Accessed 2026-02-04.

\bibitem{vctkdatashare}
{University of Edinburgh DataShare}.
\newblock Cstr vctk corpus: Data and license information (datashare record).
\newblock \url{https://datashare.ed.ac.uk/handle/10283/2119?show=full}, 2026.
\newblock Accessed 2026-02-04.

\bibitem{commonvoicecriteria}
{Mozilla}.
\newblock Common voice dataset criteria / documentation.
\newblock \url{https://commonvoice.mozilla.org/criteria}, 2026.
\newblock Accessed 2026-02-04.

\end{thebibliography}
\end{document}